\documentclass[a4paper]{spie}

\usepackage{amsmath,amsfonts,amssymb}
\usepackage{graphicx, float}
\usepackage[colorlinks=true, allcolors=blue]{hyperref}

\title{Interpretable deep learning regression for \\breast density estimation on MRI}

\author[1]{Bas H.M. van der Velden}
\author[1]{Max A.A. Ragusi}
\author[1]{Markus H.A. Janse}
\author[2]{Claudette E. Loo}
\author[1]{Kenneth G.A. Gilhuijs}
\affil[1]{Image Sciences Institute, University Medical Center Utrecht, Utrecht University, Utrecht, the Netherlands}
\affil[2]{Department of Radiology, the Netherlands Cancer Institute -- Antoni van Leeuwenhoek Hospital, Amsterdam, the Netherlands}

\begin{document} 
\maketitle

\begin{abstract}
Breast density, which is the ratio between fibroglandular tissue (FGT) and total breast volume, can be assessed qualitatively by radiologists and  quantitatively by computer  algorithms.  These  algorithms often rely on segmentation of breast and FGT volume. In this  study, we propose a method to directly assess breast density on MRI, and provide interpretations of these assessments. 

We assessed breast density in 506 patients with breast cancer using a regression convolutional neural network (CNN). The input for the CNN were slices of breast MRI of 128 $\times$ 128 voxels, and the output was a continuous density value between 0 (fatty breast) and 1 (dense breast). We used 350 patients to train the CNN, 75 for validation, and 81 for independent testing.  We investigated why the CNN came to its predicted density using Deep SHapley Additive exPlanations (SHAP).

The density predicted by the CNN on the testing set was significantly correlated with the ground truth densities (N = 81 patients, Spearman's $\rho = 0.86$, $P < 0.001$). When inspecting what the CNN based its predictions on, we found that voxels in FGT commonly had positive SHAP-values, voxels in fatty tissue commonly had negative SHAP-values, and voxels in non-breast tissue commonly had SHAP-values near zero. This means that the prediction of density is based on the structures we expect it to be based on, namely FGT and fatty tissue.

To conclude, we presented  an  interpretable  deep  learning  regression  method  for  breast  density  estimation  on  MRI  with promising results.
\end{abstract}

\let\thefootnote\relax\footnote{This paper has been published as: Van der Velden, B.H.M., Ragusi, M.A.A., Janse, M.H.A., Loo, C.E., Gilhuijs, K.G.A. ``Interpretable deep learning regression for breast density estimation on MRI.'' \textit{Medical Imaging 2020: Computer-Aided Diagnosis.} Vol. 11314. International Society for Optics and Photonics, 2020. doi: \href{https://doi.org/10.1117/12.2549003}{https://doi.org/10.1117/12.2549003}.}

\section{INTRODUCTION}
Breast density -- the ratio between fibroglandular tissue (FGT) and breast tissue -- is an important risk factor for breast cancer \cite{wolf1976,mcco2006,boyd2007}. This density can be assessed on e.g. mammography and magnetic resonance imaging (MRI), and is typically scored by a radiologist as one of four incremental categories \cite{morr2013}.

Computer algorithms have provided quantitative assessment of breast density on MRI. These algorithms often rely on identification of the breast volume by removing pectoral muscle and air, followed by segmentation of the FGT \cite{wei2004,nie2008,wu2013}. More recently, deep learning has been proposed for FGT segmentation \cite{dalm2017, ivan2019}.

Deep learning could also be used to directly assess breast density without segmentation steps. In this case, it would, however, be desirable to interpret on what basis the algorithm gave its result. In this study, we propose a method to directly assess density and provide interpretations of these assessments.

\section{Materials and methods}
We used a regression convolutional neural network (CNN) to assess breast density, and examined why the CNN came to its result using Deep SHapley Additive exPlanations (SHAP) \cite{lund2017}. The following paragraphs describe this in more detail.

\subsection{Patients}
We consecutively included 506 patients with early-stage unilateral invasive breast cancer. These patients received a preoperative T\textsubscript{1}-weighted MRI with imaging parameters: repetition time 8.1 ms, echo time 4.0 ms,  flip angle 20$^\circ$, isotropic voxel size 1.35 $\times$ 1.35 $\times$ 1.35 mm\textsuperscript{3}.

\subsection{Data preparation and ground truth creation}
For each patient, we removed field inhomogeneities using N4 biasfield correction \cite{tust2010}. We normalized the MR image between zero and one based on the 2.5\textsuperscript{th} and 97.5\textsuperscript{th} intensity percentiles and clipped intensities outside that range. We extracted a slab of 20 sagittal slices around the center of each breast -- yielding 20\,240 breast slices -- and resized them to 128 $\times$ 128 voxels.

We based the ground truth on previously generated breast and FGT segmentations \cite{veld2015}. These segmentations were manually checked in previous studies \cite{veld2015, knut2016,veld2017,veld2018,veld2019}. For each slice, we defined the density as fraction of the number of FGT voxels divided by the number of breast voxels (Figure 1).

\begin{figure} [ht!]
   \begin{center}
   \includegraphics[width=0.9\textwidth]{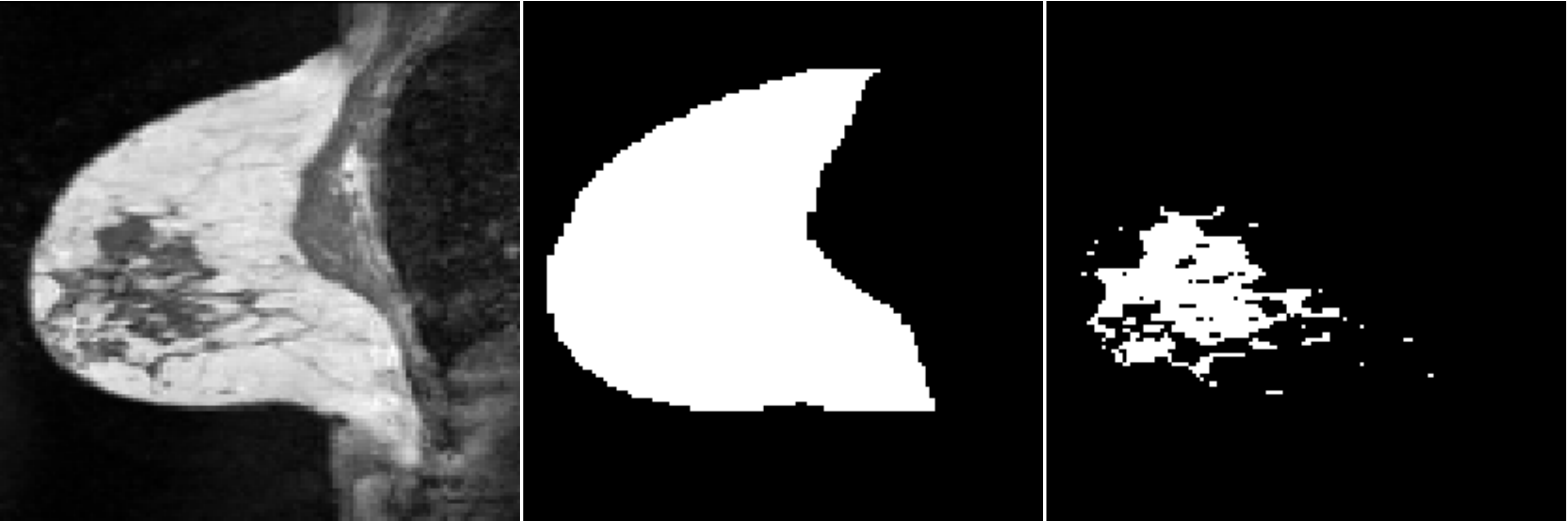}
   \end{center}
   \caption[] 
   {\label{fig:fig1} 
Example of a sagittal T\textsubscript{1}-weighted MR image slice (left) with corresponding breast segmentation (middle) and fibroglandular tissue (FGT) segmentation (right). The ground truth was created by dividing the number of voxels in the FGT segmentation by the number of voxel in the breast segmentation for each slice.}
\end{figure} 

\subsection{Regression CNN}
We used a regression convolutional neural network (CNN)\cite{vos2016} to estimate the density per slice. This CNN consisted of five convolution layers with a 3 $\times$ 3 kernel size, a 2 $\times$ 2 stride, a rectified linear unit activation, 50\% dropout, and batch normalization. These five layers were followed by two densely connected layers and an output node with a linear activation. We used the mean absolute percentage error as loss and an Adam optimizer with a learning rate of 0.001 \cite{king2014}.

We split the slices on patient-level: 14\,000 slices corresponding to 350 patients were used for training the CNN (100 epochs, mini-batches of 100 slices), 3\,000 (75 different patients) for validation, and 3\,240 (81 different patients) for independent testing. For each slice in testing, the CNN returned a density-value between 0 (fatty breast) and 1 (dense breast). The correlation of these density-values with the ground truth density was assessed using Spearman's $\rho$.

\subsection{Interpretation of CNN results}
We used Deep SHAP for interpretation of the CNN results \cite{lund2017}. Deep SHAP is a combination of DeepLIFT and SHapley Additive exPlanations \cite{shri2017,lund2017}. To assess the background signal needed for the Deep SHAP analysis, we randomly sampled 100 training slices.

For each slice, Deep SHAP yields a map of SHAP-values. Each pixel in this SHAP-values map represents the contribution of that pixel to the final decision. Hence, a higher absolute value corresponds to the pixel being more important for the prediction.

\section{Results}
\subsection{Regression CNN}
The density-values predicted by the CNN on the testing set were significantly correlated with the ground truth densities (N = 81 patients, Spearman's $\rho = 0.86$, $P < 0.001$).

\begin{figure} [ht!]
   \begin{center}
   \includegraphics[width=0.65\textwidth]{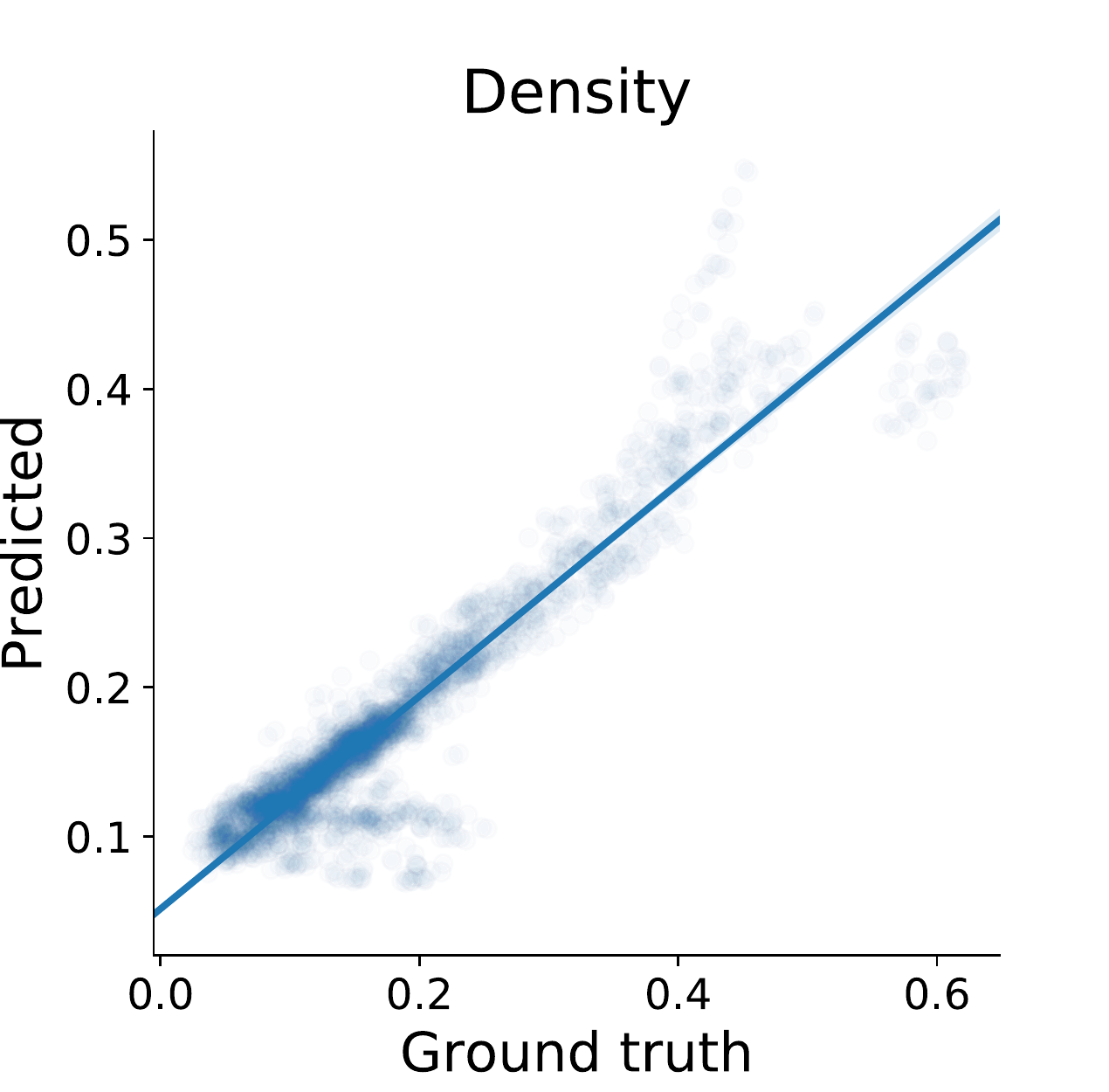}
   \end{center}
   \caption[] 
   { \label{fig:fig2} 
The density-values predicted by the regression convolutional neural network in the testing set (N = 81 patients) show a significant correlation with the ground truth densities (Spearman's $\rho = 0.86$, $P < .001$).}
\end{figure} 

\subsection{Interpretation of CNN results}
Inspection of the SHAP-value maps shows that in slices where the density predicted by the CNN matched the ground truth density, positive SHAP-values commonly occur in the glandular tissue, while negative SHAP-values occur in the fatty tissue (Figure 3). Voxels in the air, heart, or pectoral muscle are mostly ignored (Figure 3).

In slices where the predicted density deviated from the ground truth density, the SHAP-value maps are able to visualize where this deviation originated from. For example, in Figure 4B, the SHAP-value map shows overestimation in a patient with extremely dense breasts.

\begin{figure} [ht]
  \begin{center}
  \includegraphics[width=1.0\textwidth]{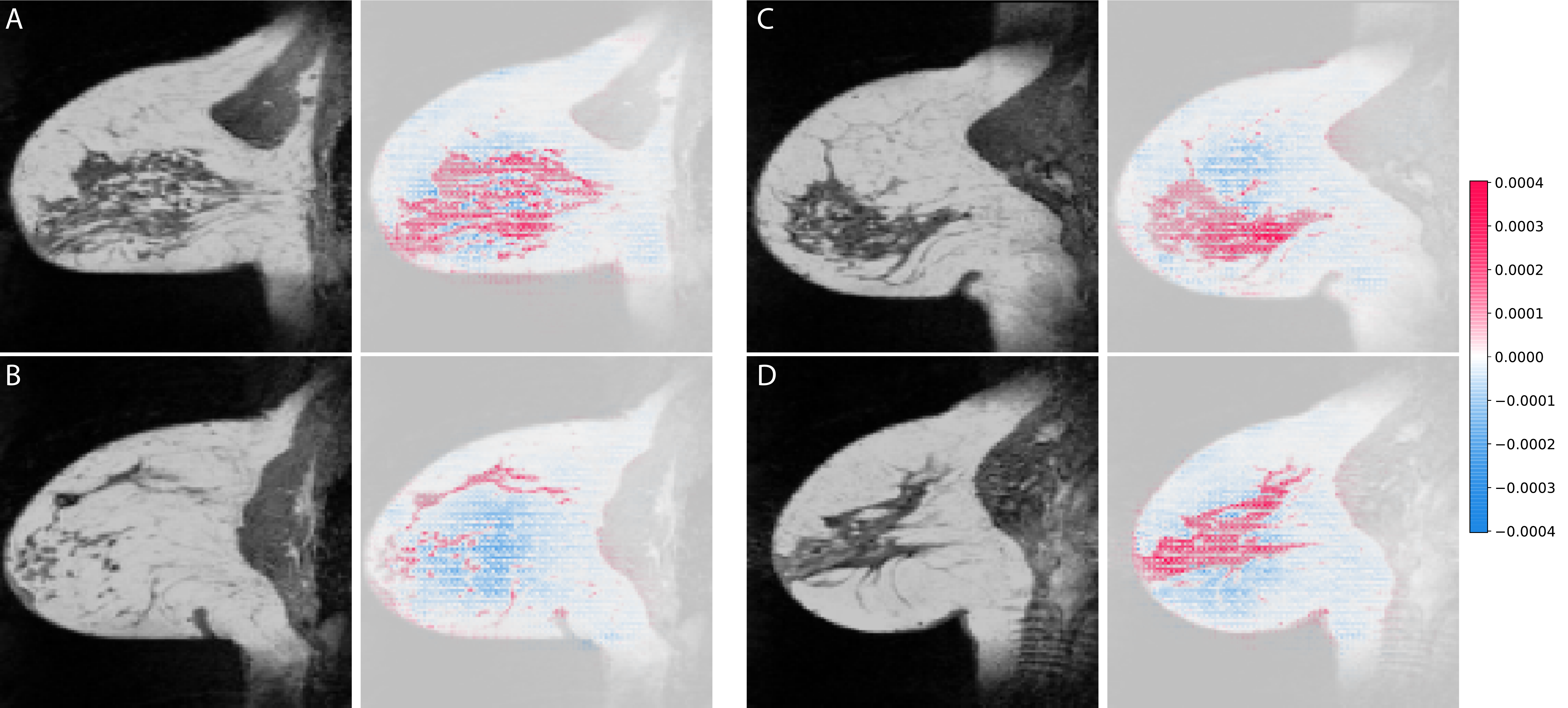}
  \end{center}
  \caption[] 
{ \label{fig:fig3} 
Four examples where the CNN correctly predicted the density of the slice: all these predictions were within 0.01 difference from the ground truth. Of each image pair, the left image is the slice the density was predicted on, the right image is a SHapley Additive exPlanations (SHAP)-map. The slice is plotted underneath with an opacity of 50\% to show anatomical information. It can be seen that positive SHAP-values (red) occur in the fibroglandular tissue, while negative SHAP-values occur in the fatty tissue. The areas of the slice that do not influence density -- such as air and pectoral muscle -- are indeed not predictive of the CNNs explanation.}
\end{figure}

\begin{figure} [ht]
  \begin{center}
  \includegraphics[width=1.0\textwidth]{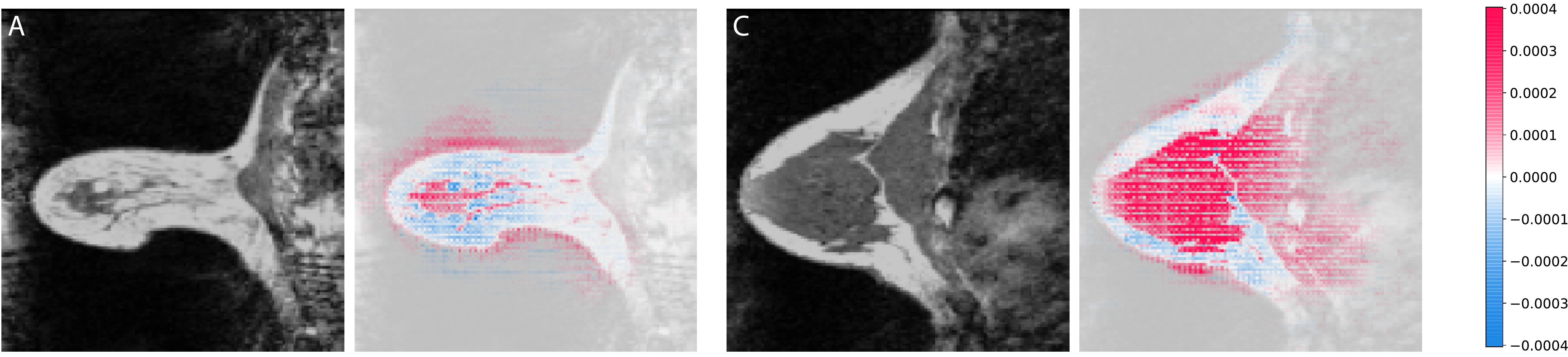}
  \end{center}
  \caption[] 
{ \label{fig:fig4} 
Two examples where the CNN overestimates the density of the slice. A: the predicted density is 0.17 and the ground truth density 0.08. In patient B, the predicted density is 0.55 and the ground truth density 0.45. These errors could be due to differences in anatomy with respect to the training set. Interpretation of the result using the SHAP-map shows uncertainty around the breast in patient A -- who had relatively small breasts -- and in the pectoral muscle in patient B -- who had extremely dense breast. The images are formatted identically to Figure 3.}
\end{figure}

\section{Discussion}
We presented a combination of deep learning regression and an interpretation method of this regression for density assessment of breast MRI. The regression method was significantly correlated with the ground truth density. 

The interpretation method supported the predictions of the CNN regression by identifying which regions of the segmentation were important. Positive SHapley Additive exPlanations (SHAP)-values commonly occurred in the fibroglandular tissue (FGT), while negative SHAP-values occurred in the fatty tissue. This is as expected: more FGT in a breast means a higher density; while the same amount of FGT in a larger breast means a lower density.

Our regression method could be used as a stand-alone solution for density assessment. It does not need intermediate steps such as segmentation to assess breast density. If a radiologist chooses to know why the method came to its result, he or she can check the interpretation using the SHAP-values. The method could in principle also be used to confirm other density assessment methods.

Our method did not always coincide with the ground truth. This mainly occurred in patients who had variations in anatomy that were not common in the training set. Future work could mitigate this by using more data.

\section{NEW OR BREAKTHROUGH WORK}
We presented an interpretable deep learning regression method for breast density estimation on MRI with promising results.

\section*{ACKNOWLEDGMENTS} This work was funded by the Dutch Cancer Society (KWF), grant number 10755.

\bibliography{main}{}
\bibliographystyle{spiebib}

\end{document}